# Creating Personalised Energy Plans: From Groups to Individuals using Fuzzy C Means Clustering

## [Extended Abstract]

**Ian Dent**[*]
Intelligent Modelling and
Analysis Research Group
School of Computer Science
University of Nottingham
Nottingham, UK, NG8 1 BB
ird@cs.nott.ac.uk

**Christian Wagner**
Intelligent Modelling and
Analysis Research Group
School of Computer Science
University of Nottingham
Nottingham, UK, NG8 1 BB
cxw@cs.nott.ac.uk

**Uwe Aickelin**
Intelligent Modelling and
Analysis Research Group
School of Computer Science
University of Nottingham
Nottingham, UK, NG8 1 BB
uwe.aickelin@nottingham.ac.uk

**Tom Rodden**
Horizon Digital Economy
Research Institute
School of Computer Science
University of Nottingham
Nottingham, UK, NG8 1 BB
tar@cs.nott.ac.uk

## ABSTRACT
Changes in the UK electricity market mean that domestic users will be required to modify their usage behaviour in order that supplies can be maintained. Clustering allows usage profiles collected at the household level to be clustered into groups and assigned a stereotypical profile which can be used to target marketing campaigns. Fuzzy C Means clustering extends this by allowing each household to be a member of many groups and hence provides the opportunity to make personalised offers to the household dependent on their degree of membership of each group. In addition, feedback can be provided on how user's changing behaviour is moving them towards more "green" or cost effective stereotypical usage.

## Categories and Subject Descriptors
H.4 [**Information Systems Applications**]: Miscellaneous

---

[*]Corresponding author

## 1. INTRODUCTION
The electricity market in the UK is being subjected to various pressures. Some of these are due to the history and current design of the National Grid and others are arising from worldwide trends, such as the need to reduce carbon emissions and the declining sources of hydro-carbon fuels. New technologies, such as electric cars needing household charging facilities, will be much more prevalent. The information available to monitor and to influence the electricity usage will grow rapidly, particularly with the roll out of Smart Meters planned for completion in the UK by 2019. In addition, the drive to change the mix of electricity generation technologies in order to reduce greenhouse gas emissions, the desire to reduce carbon dioxide by changing non-electric demand such as gas central heating to the electricity network, and the impact of climate change on altering electricity demand and the greater occurrence of extreme weather events will increase the difficulties in providing a stable and cost effective supply.

One approach to addressing these pressures is the application of Demand Side Management (DSM) techniques to achieve changes in consumer behaviour [5]. The Desimax project [3] (of which this work forms part) concentrates on applying DSM techniques to achieve overall electricity supply system efficiency.

DSM techniques requires an understanding of the spread of electricity usage of each household - the load profile. Once this is known, the most appropriate households to target and the desired changes can be determined (e.g. moving the time of peak demand).

## 2. METHODOLOGY
Clustering techniques (e.g. Kmeans) allow for the categorisation of households into a small number of clusters to which targeted marketing campaigns can be applied. Each household is assigned to one cluster but, in reality, may show aspects of many clusters. The Fuzzy C Means clustering algorithm [4] allows the allocation of households to many clusters together with an indication of the degree to which the household is a member of a cluster.

Testing has been by making use of hourly usage data collected from 93 households in Milton Keynes over a period of 2 years. The data is split into winter and summer seasons and into weekdays and weekends. Any days containing missing readings are omitted from the analysis and the data is normalised into 0:1 range and the mean calculated for each hour of the day. An average reading for each of the 93 households is obtained that consists of 24 hourly readings, each of which are calculated by averaging each day over the readings for the winter weekends.

[1] describes applying K means and Self Organised Maps algorithms to the Milton Keynes data in order to reach a set of 9 clusters where each of the households is assigned to one of the stereotypical profiles. This paper extends the work using Fuzzy C Means to allow a household to be assigned to multiple stereotypical profiles.

## 3. RESULTS
The clusters derived from the Fuzzy C means algorithm are shown at Figure 1. Each household has a degree of membership of each of the clusters as shown by the example at Figure 2 where the household is a member of 5 clusters to some degree.

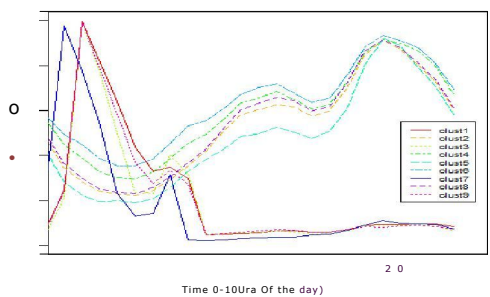

**Figure 1: Representative Clusters**

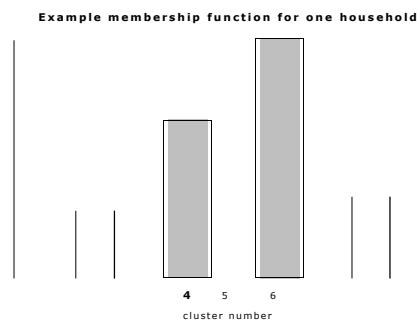

**Figure 2: Example cluster membership distribution**

## 4. CONCLUSIONS
The suggested approach allows energy suppliers to define offers for each of the nine stereotypical profiles but to offer their customers individualised tariffs based on their degree of membership of the clusters aligned with the stereotypical profiles.

As an example of how this may be used, the energy supplier may wish to see some movement of usage of electricity for the households similar to "cluster5" from the peak in early evening to the much lower usage around the middle of the day. They may develop a tariff offer for "cluster5" and then offer this to their customers with a weight depending on the household's membership of that cluster. This could be combined with offers for the other clusters such that each household is offered a personalised tariff made up of the offer for each cluster weighted by the membership of each cluster.

The results could also be used to feedback to consumers on how their day to day behaviour is changing and how they are gradually moving towards more "green" stereotypical usage patterns in the same way that dieting regimes show success by feedback of gradual short term improvements to the dieters.

## 5. ACKNOWLEDGEMENTS
The Milton Keynes data was accessed through the UK Energy Research Centre Energy Data Centre (UKERC-EDC). Our acknowledgements to the Building Research Establishment, which provided access to the original 1990 data set from Milton Keynes Energy Park, and to Bartlett School of Graduate Studies, University College London for processing and cleaning the raw data. [2] [6]

This work was possible thanks to EPSRC grant references EP/I000496/1 and EP/G065802/1.